\documentstyle[12pt]{article}
\input{psfig}
\begin{document} 
\begin{titlepage}	
\vskip 1.5cm
\begin{center}
{\Large {\bf Study of Cluster Fluctuations in
Two-dimensional q-State Potts Model$^\dagger$}} 
\end{center}
\vskip 2cm
\centerline{\bf Meral Ayd{\i}n, Yi\u{g}it G\"{u}nd\"{u}\c{c} and Tar{\i}k \c{C}elik }
\vskip 0.2cm
\centerline{\bf Hacettepe University, Physics Department,}
\centerline{\bf 06532  Beytepe, Ankara, Turkey }
\vskip 1cm

{\normalsize {\bf Abstract} The two-dimensional Potts Model with
$2$ to $10$ states is studied using a cluster algorithm to
calculate fluctuations in cluster size as well as commonly used
quantities like equilibrium averages and the histograms for
energy and the order parameter.  Results provide information
about the variation of cluster sizes depending on the
temperature and the number of states.  They also give evidence
for first-order transition when energy and the order parameter
related measurables are inconclusive on small size lattices.  }

\vskip 0.5cm

{\small {\it Keywords:} Phase transition, Potts model, clusters, Monte Carlo
simulation} 

\vskip 1.0cm

$\dagger${This project is partially supported by Turkish Scientific and Technical
Research Council (T\"{U}B\.{I}TAK) under the project TBAG-1141 and TBAG-1299.}
\end{titlepage}	
\pagebreak

\section{Introduction}

Identifying the phase structure of a statistical mechanical
system has been one of the most challenging problems of
contemporary physics.  Second-order phase transitions are well
understood in the contexts of renormalization group
\cite{Niemeijer:1976} and finite size scaling theories
~\cite{Barber:1983}. For the case of first-order phase
transitions, (for a recent review, see~\cite{Billoire:1995}) the
difficulty arises with the systems possesing finite, but very
large correlation length.  In such systems the transition can
easily be confused with second-order phase transition when the
measurements are done on a finite system. 
One of the commonly used methods to determine the order of a
phase transition is due to Challa et
al~\cite{Challa:1986}, which considers distributions of energy
or order parameter and compares these distibutions with a single
or a double gaussian for second- or first-order phase transition
cases, respectively. This method is based on the argument that
for a first-order phase transition the barrier between the
metastable states will be widened and hence distinct two
gaussians will be appearent with increasing lattice size. 
Another method employed by Lee and Kosterlitz 
~\cite{Lee:1990,Lee:1991} considers the minima in
free energy, and similar arguments are also valid for
distinguishing first-order phase transition from a second-order
one.  In both of these methods the main
idea is that the system remains in metastable states for a
considerable amount of time. Even if there exist metastable
states, the size of the system may prevent the observation of
the double-peak behavior of the energy distribution. For the
systems with finite but very large correlation length, where a
first-order phase transition resembles second-order one, one
needs to compare different size lattices in order to reach a
conclusion on the order of the transition.  The well-known
examples of these systems are 5-state Potts model in
2-dimensions~\cite{Lee:1990,Baxter:1973} and 3-state Potts model
in 3-dimensions~\cite{Lee:1991,Ohta:1993} as well as finite
temperature SU(3) lattice gauge theory~\cite{Fukugita:1989}.

$\;$

One of the major elements of a phase transition is the formation of clusters.
Observing the clusters and their variations in size 
is expected to lead  a better understanding of phase transitions
occuring in the system.
In the present work, the critical behaviour of the  two-dimensional
Potts model with $2$ to $10$ states is studied using cluster 
algorithm. Fluctuations in cluster size 
as well as  the histograms for energy and the  order parameter
are obtained as a function of the temperature for different number 
of states.

$\;$

Section II gives some information about the model and the
algorithm.  Results and discussions are given in section III,
and the conclusions are presented in section IV.

\section{Model and the method}

The Hamiltonian of the two-dimensional Potts model~\cite{Potts:1952} is given by 
\begin{equation}
     {\cal H} = K \sum_{<i,j>} \delta_{\sigma_{i},\sigma_{j}}.
\end{equation}
Here $K=J/kT$ ; where $k$ and $T$ are the Boltzmann constant and
the temperature respectively, and $J$ is the magnetic
interaction between spins $\sigma_{i}$ and $\sigma_{j}$, which
can take values $1,2, ..., q$ for the $q$-state Potts model.
The order parameter $(OP)$ can be defined through the relation
\begin{equation}
OP=\frac{q\rho^{\alpha}-1}{q-1}
\end{equation}
where $\rho^{\alpha}=N^{\alpha}/L^{D}$, $N^{\alpha}$ being the number of
spins with $\sigma=\alpha$, $L$ the linear size and $D$ the
dimensionality of the system. The average cluster size (CS) is calculated by taking
the average over the number of clusters ($N_c$),  
\begin{equation}
CS = \displaystyle{{1}\over{N_c}}< \sum_{i=1}^{N_c} C_i >  
\end{equation}
where $C_{i}$ is the number of spins in the ${\rm i^{th}}$ cluster.
The two-dimensional Potts model undergoes a second-order 
phase transition for $2,3$ and $4$ states, while a first-order phase 
transition is known to occur for higher number of states\cite{Baxter:1973}.
Reader can refer to the review article by Wu~\cite{Wu:1982}
for detailed information about the model.     

$\;$

First-order transitions can be recognized by studying
double-peak behaviour of the energy probability distribution
function $P(E)$~\cite{Challa:1986}. When the transition is
first-order, order parameter probability distribution function 
$P(OP)$ should also exhibit the same behaviour.  If the energy
distribution, measured on sufficiently large lattices, is a
single gaussian at the critical point, this is a clear
indication of a second-order phase transition, and having two
gaussian distributions is an indication of a first-order
transition as mentioned above. Even if the energy has a single
gaussian distribution, the order parameter or the cluster size
distribution may mimic a first-order transition. Larger lattices
may lead to double gaussian shape of the energy distribution but
measurements made on moderate size lattices may be inconclusive.
Moreover, a double-peak behaviour in the distribution of the
order parameter may be misleading~\cite{Eisenriegler:1987}.

$\;$

Another observable which can provide information about critical
behavior is the average cluster size (CS).
The study of variations in cluster size for 
different $K$ and $q$ should  give some indications about
the nature of the transitions occuring in the system.
By these considerations, identification of the order of phase
transition is reduced to correct identification of the various size
clusters which requires an extra computational effort when a usual
local updating algorithm is used. The reduction of critical slowing
down by a cluster flip algorithm was first introduced by Swendsen and
Wang~\cite{Swendsen:1987} and later modified by Wolf~\cite{Wolf:1989}.
The cluster algorithm used in this work is the same as Wolf's
algorithm, with the exception that, before calculating the
observables, searching the clusters is continued until the total
number of sites in all searched clusters is equal to or exceeds the
total number of sites in the lattice.

\section{Results and discussions}

After thermalization with $10^{4} - 5 \times 10^{4}$ sweeps, $5
\times 10^5$ iterations are performed for $32 \times 32$ and $64
\times 64$ lattices of the $2$ to $10$-state models at different
values of the coupling $K$. Majority of the runs are started
using a disordered initial configuration, and the averages
calculated for both ordered and disordered starting
configurations are observed to be equal within statistical
errors.  Longer runs with up to $10^{6}$ iterations are done
near the finite-size critical value $K_{c}$ at each state
$q$. As a result of these runs, equilibrium averages for energy,
cluster size and the order parameter are obtained, specific heat
and the Binder cumulant are calculated as a function of $K$ for
each value of $q$.  These quantities are used as a test ground
for the algorithm used as well as for the thermalization and the
accuracy, and they show expected behavior for different lattice
sizes and different number of states, therefore they will not be
reported here.  $K_{c}$ values, which are estimated at the peak
values of the specific heat for the $64 \times 64$ lattice are
close to $K^{\infty}_{c}$ values for the infinite lattice within
an error of about one percent.  The errors calculated
using Jacknife analysis are small (less than one percent
for most of the quantities, within about $2$ percent for the
specific heat) and they can be decreased by increasing the
number of Monte Carlo iterations.

%$\;$

The histograms for energy, order parameter, and the average
cluster size  as a function of the number of iterations 
are obtained for $q=2$ to $10$ near $K_{c}$. 
The models with $q=2$ and $3$ have similar histograms with a single gaussian, and
for the cases $q \ge 5$, all
histograms have distinct double peaks,
getting higher and sharper as $q$ increases.
In the histograms for energy, double peaks
occur for $q \ge 5$ with almost equal maxima
near the $K_{c}$ values estimated using the specific heat data, as
expected.
Figure 1 shows the histograms
for energy and the order parameter for $q=3, 4$ and $5$ on $64 \times 64$ 
lattice obtained at $K_{c}$. Plots for $q=3$ have single peaks for both 
energy and the order parameter. The histograms for $q=5$ show indication 
of the first-order character, with the appearance of double peaks in both plots. 
The energy histogram for $q=4$ is similar to the one for $q=3$, but the 
order parameter histogram shows a variation which looks more similar to 
the plot for $q=5$. 

$\;$

The fluctuations $FCS$ ($FCS=<(CS)^{2}>-<CS>^{2}$)
in the average cluster size $CS$ are calculated
for different values of $q$ on $64 \times 64$ lattice as a function 
of $K$. Figure 2 shows $FCS$ for $q$ varying from $3$ to $7$
and the $CPU$ times for the corresponding Monte Carlo runs.  
As can be seen from these plots, $FCS$ and the $CPU$ time give similar
information about the formation of the clusters
$(1/FCS$ has a similar variation as the $CPU$ time).
When $K$ is small (i.e. the temperature is high), the cluster sizes are
small, hence $FCS$ is small, and the algorithm spends a long
time to search for the clusters small in size but large in number. As $K$
increases, the cluster sizes are getting larger, so the $CPU$ time decreases, 
but $FCS$ increases due to the existence of the small-size clusters
as well as the large ones. At the critical point, formation of the
largest clusters results in the largest value of $FCS$, with
the corresponding minimum value of the $CPU$ time, since there are only
a few clusters to be searched. One can make a similar discussion
when the critical point is approached from above  
(in the region where $K>K_{c}$), 
with the exception that the $CPU$ time is almost constant for large $K$,
as mentioned in section II.

$\;$

$FCS$ plots also show that the maximum value of $FCS$  
increases as $q$ increases. When $q$ is getting larger, decreasing
correlation length (due to increasing first-order character) results
in breaking large clusters into smaller ones. Near the critical point,
the tendency of forming large clusters together with the disintegration
process causes $FCS$ to increase with increasing $q$. 
In the second-order phase transitions, long range correlations keep 
the clusters as they are for a very long time or for large number of 
iterations (hence $FCS$ is small for small $q$). 
The $FCS$ and $CPU$ plots show that the models with $q > 4$
behave differently at two sides of the critical point. 
The plots for $q=3$ are in the form of
curves with a smooth variation, passing through a minimum near $K_{c}$.
Plots for the other cases show different behaviour in two separate regions around
the critical point, with a sharp variation near $K_{c}$ value. The $CPU$ data 
for $q=3$ can be fitted to a polynomial function around $K_c$ 
at both sides of the critical point, but no fit seems to be possible for
$q > 4$. The only possibility is the fit of two distinct 
polynomials to two separate curves, for $K<K_{c}$ and $K>K_{c}$.
 
$\;$

%\pagebreak    

\section{Conclusions}

The two-dimensional Potts model with  
$2$ to $10$-states is studied using cluster algorithm 
for lattices with sizes $32 \times 32$ and $64 \times 64$.
Energy, order parameter, cluster size, fluctuations in 
the cluster size, specific heat and the Binder cumulant are calculated, 
and the histograms for energy, order parameter and the cluster size are
obtained as a function of the coupling $K$.  

$\;$

Studies in this work are focused on the cluster size related observables. 
One of these observables is the fluctuations in the 
cluster size (FCS) which measures the extent of the coexistence of different 
size clusters. This operator is very sensitive to the correlation length 
and exhibits sharp peak at the transition region 
if the transition is first-order. The similar information can 
also be obtained from the CPU time spent to calculate the averages for 
a certain number of iterations. If there exist only small clusters, due 
to the modified algorithm used in this work, the CPU time is extremely 
large and growing cluster size reduces this time. In the first-order 
case, small clusters grow very fast near $K_{c}$, while in the second-order case, 
the average cluster size grows smoothly.  
Both figures 2.a (FCS) and 2.b (CPU) indicate 
that on $64 \times 64$ lattice $q= 3\, {\rm and}\, 4$ have 
different character than  $q \ge 5$ Potts models. 
From these two figures one can conclude that for 
the systems with varying correlation length, $FCS$ and $CPU$ are sensitive 
to the correlations even though the correlation length is much larger 
than the system size.

$\;$
 
Further work is planned, using the same algorithm,
to study cluster distributions in detail as well as
random-field and random-bond Potts models.

\vskip 0.5cm

\section*{Acknowledgements}  

Authors are indepted to Prof. D. Stauffer for his valuable comments.

\pagebreak

\pagebreak

\section*{Figure captions}

\begin{description}

\item {Figure 1.} a) Energy, b) the order parameter histograms,   
as a function of the number of iterations for $q = 3, 4$ and $5$
on $64 \times 64$ lattice near $K_{c}$.

\item {Figure 2.} a) Fluctuations in the average cluster size, b) the
CPU time, as a function of $K$ for $q$ varying from $3$ to $7$ on $64 \times 64$ 
lattice. Computations are done using  IBM-RS/6000 workstations.   
The errorbars in $FCS$ plots are about the size of the data points and are visible only
for large $q$.

\end{description}

\pagebreak

\begin{figure}
\psfig{figure=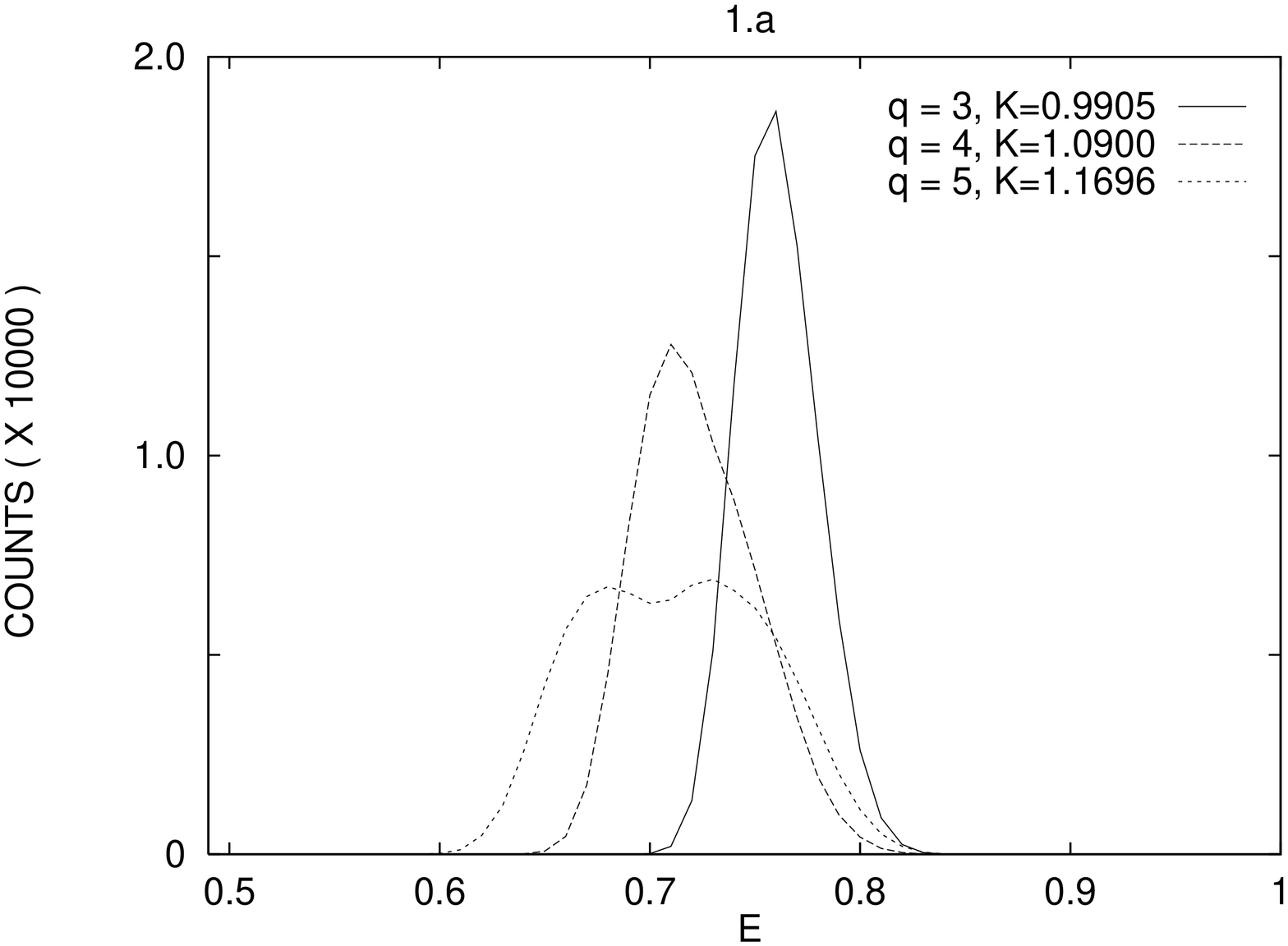,height=8cm,width=12cm,angle=-90}
\vskip 1cm
\psfig{figure=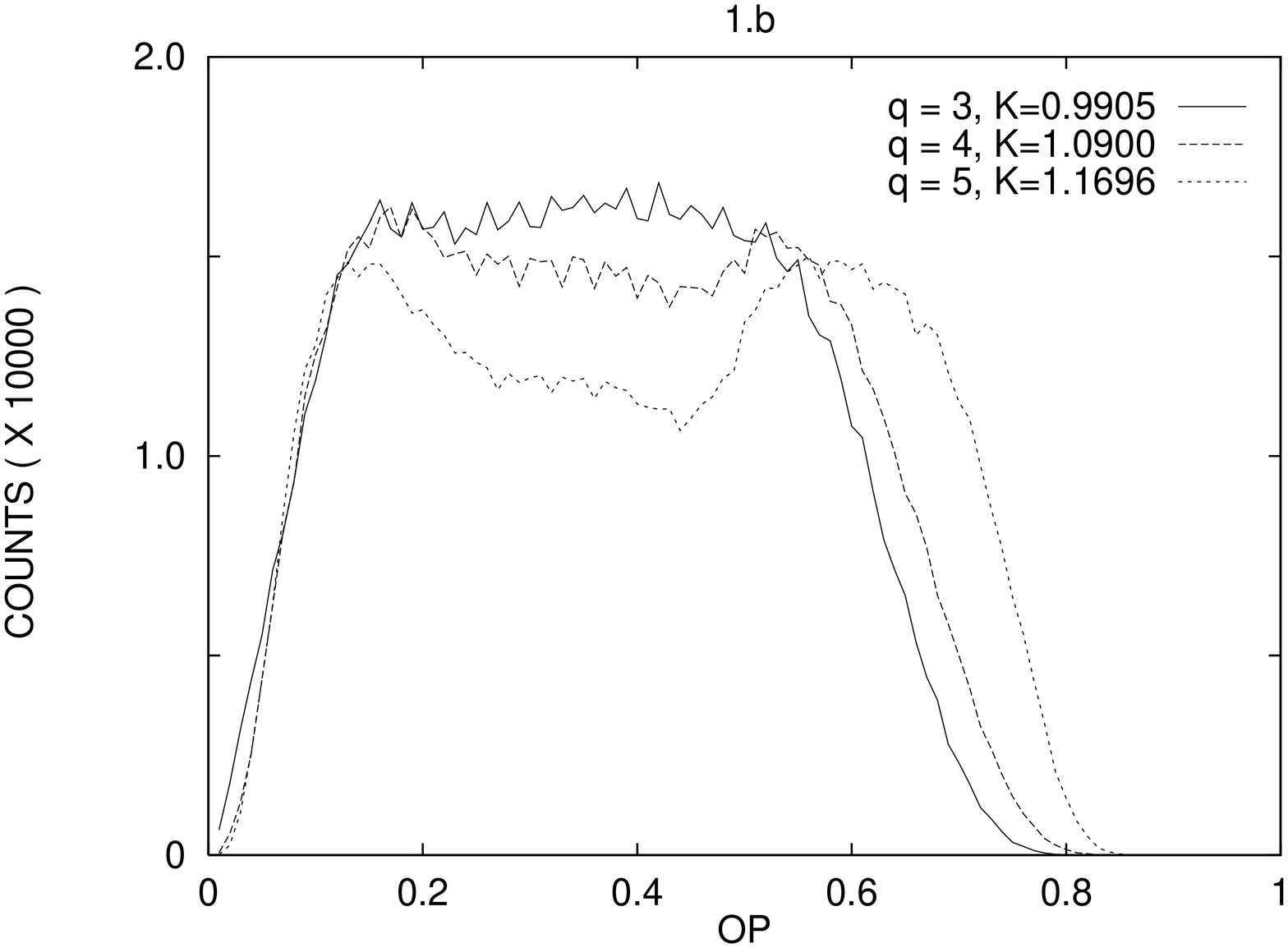,height=8cm,width=12cm,angle=-90}
\caption{}
\end{figure}
\begin{figure}
\psfig{figure=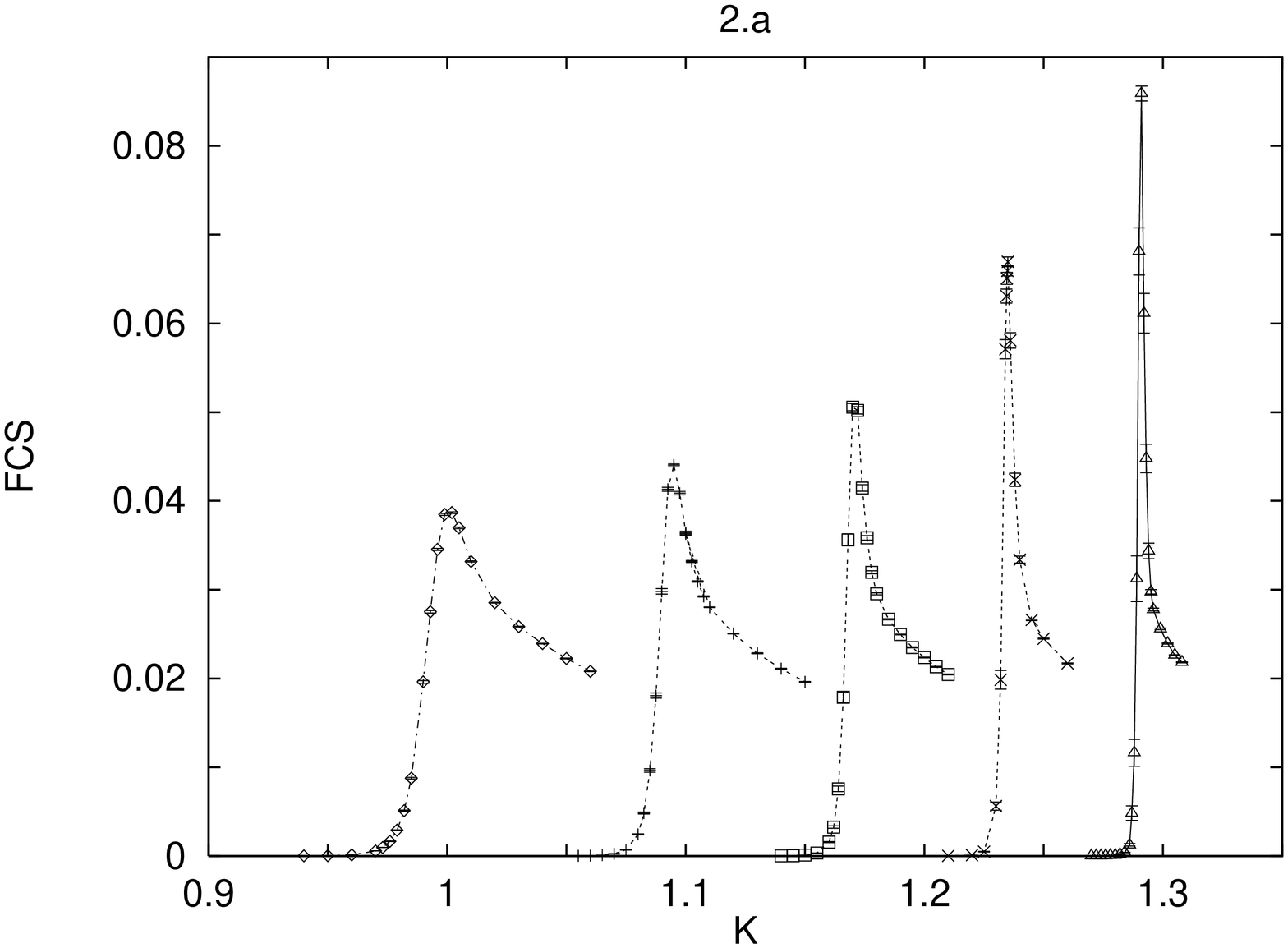,height=8cm,width=12cm,angle=-90}
\vskip 1cm
\psfig{figure=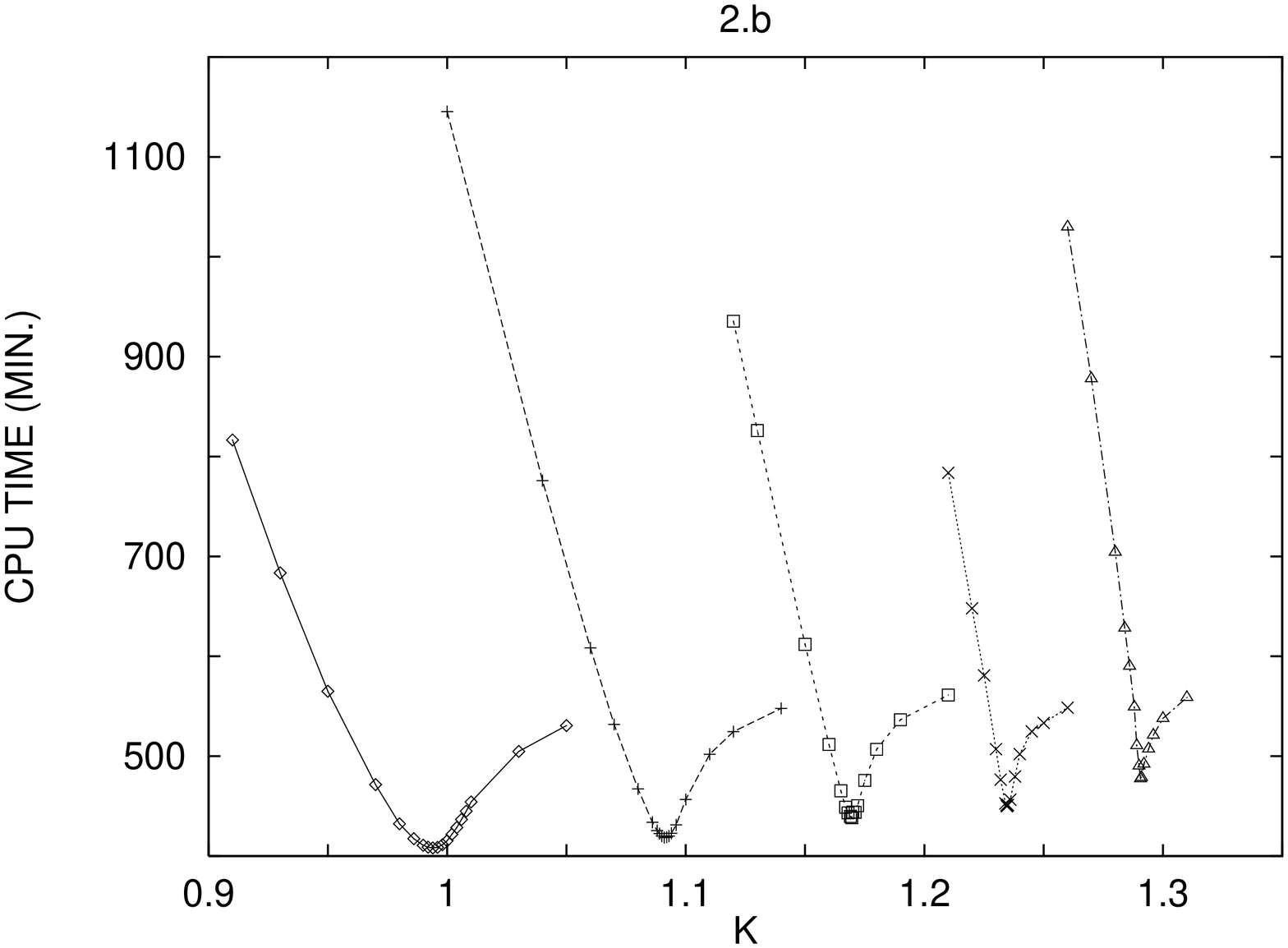,height=8cm,width=12cm,angle=-90}
\caption{}
\end{figure}
\end{document}